\documentclass[3p]{elsarticle}

\usepackage{lineno,hyperref,amsmath}
\modulolinenumbers[5]
\usepackage{amssymb,color}
 \usepackage{graphicx}

\def\const{\mbox{const\;}}

\journal{Journal of \LaTeX\ Templates}

\begin{document}

\begin{frontmatter}

\title{Adiabatic Interactions of Manakov Solitons -- Effects of Cross-modulation}

\author{V. S. Gerdjikov$^{1,2,4}$, M. D. Todorov$^{*\,3}$, A. V. Kyuldjiev$^2$}
\address{$^1$Institute of Mathematics and Informatics, Bulgarian Academy of Sciences,
8 Acad. G. Bonchev str.,  1113 Sofia, Bulgaria \\
$^2$Institute for Nuclear Research and Nuclear Energy, Bulgarian
Academy of Sciences, 72 Tzarigradsko Chaussee, Blvd., 1784 Sofia, Bulgaria \\
$^3$Dept of Diff. Equations, Faculty of Applied
Math. and Informatics,  Technical University of Sofia, 8 Kl. Ohridski Blvd., 1000 Sofia, Bulgaria\\
$^4$ Institute for Advanced Physical Studies, New Bulgarian University, 21 Montevideo Street, Sofia 1618, Bulgaria}
\ead{gerjikov@inrne.bas.bg, mtod@tu-sofia.bg, kyuljiev@inrne.bas.bg}


\cortext[mycorrespondingauthor]{Corresponding author}


\begin{abstract}
We investigate the  asymptotic behavior of the Manakov  soliton trains perturbed by cross-modulation
in the adiabatic approximation.  The multisoliton interactions in the adiabatic approximation are modeled by a generalized Complex Toda chain
(GCTC).  The cross-modulation  requires special treating for the
evolution of the polarization vectors of the solitons. The numerical predictions of the Manakov system are compared with the perturbed GCTC.
For certain set of initial  parameters GCTC describes very well  the long-time evolution of the Manakov soliton trains.

\end{abstract}


\end{frontmatter}

\section{Introduction}

The Manakov model (MM) \cite{Man}
\begin{equation}\label{eq:vnls}
i\frac{\partial \vec{u}}{\partial t} + \frac{1}{2} \frac{\partial^2\vec{u}}{\partial x^2} +\langle\vec{u}{\;}^\dag, \vec{u}\rangle \vec{u}(x,t)= 0,\end{equation}
where $\vec{u}(x,t) = (u_1(x,t), u_2(x,t))^T$, $t$ -- time, $x$ -- spatial coordinate, $\vec{u}^\dag = (u_1^*,u_2^*)^T$ -- hermitian conjugate to
$\vec{u}$, $\langle\vec{u}^\dag,\vec{u}\rangle$ -- scalar product of $\vec{u}^\dag$ and $\vec{u}$,
is the first generalization of the  famous nonlinear Schr\"odinger equation \cite{ZaSh} to multi-components. It
finds a number of applications to physics: in nonlinear optics \cite{BEG,LK,Man,Y}, in Bose-Einstein condensates (BEC) \cite{Ho,IMW04,FCR,Pita,OM,VPG},
in plasma physics   and others \cite{Mon1,NMPZ}).
 We specially  mention some of them \cite{BEG,Boris,OM}, which
are closer to the types of interactions we consider.

Therefore it is an important problem to study the soliton interaction of (\ref{eq:vnls}) in the adiabatic approximation for
multicomponent nonlinear Schr\"odinger equations.
This task has been started some time ago in \cite{Thes}, \cite{G*11b}--\cite{PhD}, \cite{KS,Boris,SD}. First, it was proved that a generalized version of the complex
Toda chain (CTC) \cite{Thes,GKDM,GTK-1} describes rather well the $N$-soliton train behavior for wide set of soliton parameters.
Since this generalized CTC is also integrable it allows one also to predict the asymptotic behavior of the $N$-soliton train
for $t\to\infty$. The numerical tests comparing the trajectories of the solitons obtained as numerical solution of (\ref{eq:vnls})
with the trajectories predicted by the CTC showed an excellent agreement for up to 9-soliton trains.

The perturbed MM also finds a number of applications. For example,  the equation
\begin{equation}\label{eq:vnls2}
i\frac{\partial\vec{u}}{\partial t} + \frac{1}{2}\frac{\partial^2 \vec{u}}{\partial x^2} +\langle\vec{u}{\;}^\dag, \vec{u}\rangle \vec{u}(x,t)= V(x)\vec{u}(x,t),
\end{equation}
\noindent where $V(x)$ is an external potential can be viewed as one-dimensional model of the Gross-Pitaevski equations.\noindent
Therefore it could be used to model quasi-one-dimensional BEC \cite{FCR,Pita,VPG}. The system (\ref{eq:vnls2}) is not integrable.
However  a perturbed GCTC was derived for several special types of potentials \cite{G*11b,GKDM,GTK-1,GT-3}.

Next we consider MM with cross-modulation  parameter $\alpha_2$
\begin{equation}\label{eq:Man1}\begin{split}
i \frac{\partial u_1}{ \partial t }  + \frac{1}{2} \frac{\partial^2 u_1 }{ \partial x^2 } +
\left[ (1-\alpha_2)|u_1|^2 + (1+\alpha_2)|u_2|^2 \right]  u_1(x,t) = 0, \\
i \frac{\partial u_2}{ \partial t }  + \frac{1}{2} \frac{\partial^2 u_2 }{ \partial x^2 } +
\left[ (1+\alpha_2)|u_1|^2 + (1-\alpha_2)|u_2|^2 \right]  u_2(x,t) = 0,
\end{split}\end{equation}
Obviously for small $\alpha_2$ the above equation can be still considered as a perturbed nonlinear vector Schr\"odinger equation.
Eq.~(\ref{eq:Man1}), as other perturbed versions of MM is also non-integrable \cite{ZahSch}. Nevertheless for small $\alpha_2$ the adiabaticity is still possible
and a respective generalized CTC model can be built. This is the main goal of the investigation in this paper.
The model, Eq.~(\ref{eq:Man1}) has been  analyzed in detail previously by numerical methods in \cite{CDM,CT06,CT08,CT08'} for wide range of $\alpha_2$.

In  Sections 2 and 3 we introduce some preliminary facts and notations and apply the variational approach to the system (\ref{eq:Man1}) to derive a perturbed
GCTC model that models the $N$-soliton trains in the presence of cross-modulation \cite{G*11a,TMF} . The main
difficulty here is to derive an adequate equation for the evolution of the polarization vectors $\vec{n}_k$.
In Section 4 we briefly discuss how the GCTC can be used to investigate the asymptotic regimes of the soliton trains.
In Section 5 we  demonstrate that for some sets of soliton parameters the perturbed GCTC
gives very good description for the cross-modulated MM.
 To this end we solve the cross-modulated MM numerically by using an implicit scheme of  Crank-Nicolson type in
 complex arithmetic for the linear part of the operator and internal iterations for the nonlinear one.
 The concept of the internal iterations is applied (see \cite{CDM}) in order to ensure the implementation of the conservation laws
 on difference level within the round-off error of the calculations \cite{CT06,CT08,CT08'}.
The solutions of the relevant GCTC
have been obtained using Maple. Knowing the numeric solution $\vec{u}$ of the perturbed MM we calculate  he maxima of
$(\vec{u}\,^\dag, \vec{u})$, compare them with the (numeric solutions) for $\xi_k(t)$ of the GCTC and plot the
predicted by both models trajectories for each of the  solitons. Thus we are able to analyze the effects of the
cross-modulation on the soliton interactions.
We end with some conclusions and discussions.

\section{Posing the Problem. Adiabatic Approximation}

Here  we briefly remind the derivation of the CTC as a model describing the $N$-soliton interactions
of multicomponent NLS systems using the variational approach \cite{LisAnd,LisAnd1}.

The main idea of the adiabatic approximation to soliton interactions is as follows \cite{KS}. Consider the
MM, or one of its perturbed versions and analyze the dynamics of its solution with initial condition:
\begin{equation}\label{eq:Nstr}\begin{split}
&\vec{u}(x,t=0) = \sum_{k=1}^N \vec{u}_k(x,t=0), \\ \vec{u}_k(x,t) &= u_k(x,t)
 \vec{n}_k , \qquad u_k(x,t) = {2\nu_k {\rm e}^{i\phi_k}\over \cosh(z_k)}
\end{split}\end{equation}
where
\begin{equation}\label{eq:Nstr2}\begin{aligned}
z_k &= 2\nu_k (x-\xi_k(t)),  &\qquad  \xi_k(t)  &=2\mu_k t +\xi_{k,0}, \\
\phi_k &= {\mu_k \over \nu_k} z_k + \delta_k(t),  &\qquad  \delta_k(t) &=2(\mu_k^2+\nu_k^2) t +\delta_{k,0}.
\end{aligned}\end{equation}
The  $2$-component polarization vector is parametrized by
\begin{equation}\label{eq:nk0}\begin{split}
 \vec{n}_k = \left( \cos(\theta_k) e^{i \beta_{k}}, \sin(\theta_k) e^{-i \beta_{k}}\right)^T .
\end{split}\end{equation}
It is obviously normalized by the conditions $( \vec{n}{\,}^\dag_k , \vec{n}_k)=1 $.
The adiabatic approximation holds true if the soliton parameters  satisfy \cite{KS}:
\begin{eqnarray}\label{eq:ad-ap}
&& |\nu _k-\nu _0| \ll \nu _0, \qquad |\mu _k-\mu _0| \ll \mu _0,
\qquad |\nu _k-\nu _0| |\xi_{k+1,0}-\xi_{k,0}| \gg 1,
\end{eqnarray}
for all $k$, where $\nu _0 = {1  \over N }\sum_{k=1}^{N}\nu _k$, and $ \mu _0 =
{1 \over N }\sum_{k=1}^{N}\mu _k$ are the average amplitude and
velocity, respectively. In fact we have two different scales:
\begin{equation}\label{eq:7'}\begin{split}
 |\nu _k-\nu _0| \simeq \varepsilon_0^{1/2}, \qquad |\mu _k-\mu _0| \simeq \varepsilon_0^{1/2}, \qquad
|\xi_{k+1,0}-\xi_{k,0}| \simeq \varepsilon_0^{-1}.
 \end{split}\end{equation}


Next the basic idea of the adiabatic approximation is to derive a dynamical system for the soliton
parameters which would describe their interaction. The initial condition (\ref{eq:Nstr}) is not an exact $N$-soliton
solution evaluated at $t=0$. It involves some small ($\simeq 1\%$) contribution of radiation due to the continuous
spectrum. Adiabaticity also means that the solitons never overlap strongly. If this happens the approximation
breaks down.

Initially this idea was proposed by Karpman and Solov'ev \cite{KS}. They inserted
the initial condition in the NLS and after tedious calculations derived the dynamical system for the 8 parameters
of the  two solitons. A slightly different approach was proposed in \cite{LaHaus,Jan}. There the authors multiply the NLS eq.
by a set of orthogonal functions and integrate. Such approach requires that the set orthogonal functions is convenient and
complete.

An alternative derivation known as the variational approach was proposed
by Anderson and Lisak \cite{LisAnd,LisAnd1}, see also \cite{Boris}. Later this idea was generalized to $N$-soliton interactions
\cite{PRE*55,PRL,PLA-241,PhD} and the corresponding dynamical system for the $4N$-soliton parameters was identified as a
$N$-site CTC. The fact that the CTC, (just like its real counterpart -- the Toda chain (RTC)) is completely
integrable gives additional possibilities. A detailed comparative analysis between the solutions of the RTC and CTC
\cite{JPA1} shows that the CTC allows for a variety of asymptotic regimes, see Section 4 below. More precisely, knowing the
initial soliton parameters one can effectively predict the asymptotic regime of the soliton train. Another possible use
of the same fact is, that one can describe the sets of soliton parameters responsible for each of the asymptotic regimes.
Another important advantage of the adiabatic approach is, that one may consider the effects of various
perturbations on the soliton interactions \cite{KS,PRE*55}.

These results were extended  to treat the soliton interactions of the Manakov solitons.
We derive a generalized version of the CTC as a model describing the behavior of the $N$-soliton trains of
the MM \cite{Thes,G*11a,TMF,GKDM,GT-3}. This generalized CTC includes also the evolution of the polarization vectors $\vec{n}_k$.
Using it one can predict the asymptotic regimes of the Manakov solitons and can describe the sets of soliton parameters that are responsible for
each of the asymptotic regimes. Of course, just like for the scalar case, one can also analyze the effects of the
various perturbations on the soliton interactions.

\section{Derivation of the CTC as a model for the soliton interaction of the cross-modulated MM systems}

We start with the  Lagrangian of the cross-modulated MM systems:
\begin{equation}\label{eq:120.2}
\mathcal{L}= \int_{-\infty}^{\infty} dt\;  {i\over 2} \left[
\langle\vec{u}_t^\dag , \vec{u}\rangle- \langle\vec{u}\, ^\dag ,\vec{u}_t\rangle \right] - H.
\end{equation}
where the  Hamiltonian for the cross-modulated MM is
\begin{equation}\label{eq:Hcm}\begin{split}
H = \int_{ -\infty}^{\infty} dx \; \left( \frac{1}{2} \langle\vec{u}_x^\dag , \vec{u}_x\rangle - \frac{1}{2} (|u_1|^2 + |u_2|^2)^2 +
\frac{\alpha_2}{2} (|u_1|^2 - |u_2|^2)^2 \right)
\end{split}\end{equation}
It is easy to check that the Lagrangian equations of motion
\begin{equation}\label{eq:Lagr}\begin{split}
 \frac{\partial }{ \partial t} \frac{ \delta \mathcal{L}}{\delta u_{k,t}^*} -\frac{ \delta \mathcal{L}}{\delta u_{k}^*}=0,
 \qquad k=1,2
\end{split}\end{equation}
coincide with the equations (\ref{eq:Man1}).


The idea of the variational approach of \cite{LisAnd,LisAnd1} is to insert the anzatz (\ref{eq:Nstr}) into the Lagrangian,
perform the integration over $x$ and retain only terms of the orders of $\varepsilon_0^{1/2}$ and $\varepsilon_0$.
The first obvious observation is that only the nearest neighbors solitons will contribute such terms and
\begin{equation}\label{eq:L}
\mathcal{L}_{\rm eff} = \sum_{k=1}^{N} \mathcal{L}_k + \sum_{k=1}^{N} \sum_{n=k\pm 1} \widetilde{\mathcal{L}}_{kn}.
\end{equation}

The leading order terms are the ones in $\mathcal{L}_k$ which correspond to
 the terms involving only the $k$-th soliton (see \cite{TMF}):
\begin{equation}\label{eq:123.2}\begin{split}
\mathcal{L}_k &= 4\nu _k \left( {i  \over 2 } \left( \langle\vec{n}_k ^\dag,\vec{n}_{k,t}\rangle
-\langle\vec{n}_{k,t}^\dag,\vec{n}_k\rangle \right) +2\mu _k{d\xi_k  \over dt } - {d\delta _k  \over dt }
-2\mu _k^2 + { 2\nu _k^2 \over 3 }\right)  -  \frac{4\alpha_2 \nu_k}{3} \left(|n_{1,k}|^2 - |n_{2,k}|^2 \right)^2.
\end{split}\end{equation}
Note that the terms that do  not contain $t$-derivatives are of the order of 1. The order of the terms that contain
$t$-derivatives can be established after one derives the evolution equations. 

The second type of terms are the ones that involve only two different solitons, say $k$ and $p$. For example,
consider
\begin{equation}\label{eq:Ikp}\begin{split}
\mathcal{I}_{kp} &=\int_{ -\infty}^{\infty} dx \; \left( u_k (x,t) u_{p}^*(x,t) + u_k^* (x,t) u_{p}(x,t)\right)
\simeq  4\nu_k\nu_{p} \int_{-\infty}^{\infty} \frac{dz_k\;  4\nu_0 \cos( \phi_k - \phi_p) }{\cosh(z_k) \cosh(z_{p}) } \\
&\simeq 4\nu_0 \cos(\delta_k -\delta_p)  \Delta_{k,p} e^{-|\Delta_{k,p}|} , \qquad \Delta_{k,p} = \xi_k - \xi_p.
\end{split}\end{equation}
For $p=k\pm 1$ the right hand side of eq. (\ref{eq:Ikp}) is the largest and is of the order of $\varepsilon$:
\begin{equation}\label{eq:eps}\begin{split}
\varepsilon &\simeq \int_{ -\infty}^{\infty} dx \; |u_k (x,t) u_{k+1}(x,t)|  \simeq 2  \Delta e^{-|\Delta|}, \qquad \Delta\equiv \Delta_{k,k \pm 1}.
\end{split}\end{equation}
In estimating the integral in (\ref{eq:Ikp}) and (\ref{eq:eps}) we also did several approximations:
a) we replaced $\nu_k$ and $\nu_{k+1}$ by $\nu_0 = 1/N \sum_{k=1}^{N}\nu_k$ thus neglecting terms of the order
of $\varepsilon^{3/2}$; b) we replaced $\sinh^{-1}(\Delta)$ by $2s_{k,k+1}e^{-\Delta}$ thus neglecting terms of the order
of $\varepsilon^{2}$.

Comparing eqs. (\ref{eq:Ikp}) and (\ref{eq:eps}) we find that $\mathcal{I}_{kp} \simeq \varepsilon^{|k-p|}$. This means that
the adiabatic approximation takes into account only the nearest neighbor interactions (i.e. the ones with $p=k\pm 1$) and
neglects the ones with $|k-p| \geq 2$.

The third types of  terms  contain the effect of interactions of three and more different solitons.
It is easy to see that these integrals  contribute terms of order $\varepsilon^2$ and $\varepsilon^3$ respectively and
therefore they are also neglected.

Keeping only the  terms up to the order of $\varepsilon$ we obtain:
\begin{equation}\label{eq:Lkn}\begin{aligned}
 \mathcal{L}_{kn} &= 16\nu _0^3 e^{-\Delta_{kn}} (R_{kn}+R_{kn}^*)+\mathcal{O}(\varepsilon ^{3/2}), &\qquad
R_{kn} &= s_{kn}e^{i(\widetilde{\delta} _n- \widetilde{\delta} _k)} \langle\vec{n}_k^\dag , \vec{n}_n\rangle, \\
 \widetilde{\delta}_k &=\delta _k - 2\mu _0\xi_k, &\qquad  \Delta _{kn} &=2s_{kn}\nu _0(\xi_k- \xi_n),
\end{aligned}\end{equation}
where  $n=k\pm 1$ and $s_{k,k+1}= 1$ and $s_{k,k-1}= -1$.

The next step is to consider the effective Lagrangian $\mathcal{L}_{\rm eff}$ (\ref{eq:L}) as a Lagrangian of the dynamical system,
describing the motion of the $N$-soliton train and providing the equations of motion for the
$(2s+2)N$ ($6N$ for the Manakov case) soliton parameters.

Let us first consider the unperturbed case, \emph{i.e.}, $\alpha_2=0$. Thus  we arrive at the following set
of dynamical equations for the soliton parameters:
\begin{equation}\label{eq:139.2}\begin{aligned}
 {d\xi_k  \over dt } &= 2\mu _k ,   &\; {d\delta _k  \over dt } &=  2\mu _k^2 +2\nu_k^2 , \\
 {d\nu _k  \over dt } &= 8\nu _0^3 \sum_{n=k\pm 1}^{} e^{-\Delta _{kn}} i(R_{kn}-R_{kn}^*),   &\;
{d\mu _k  \over dt } &= -8\nu _0^3 \sum_{n =k\pm 1}^{} e^{-\Delta _{kn}} (R_{kn}+R_{kn}^*).
\end{aligned}\end{equation}
Let us briefly discuss the order of the terms in the right hand sides of eqs. (\ref{eq:139.2}). Since we have assumed that
the average velocity of the soliton train is $\mu_0=0$ then from eq. (\ref{eq:7'}) it follows that the r.h.side of the first equation
is $\mu_k \simeq \mathcal{O}(\varepsilon^{1/2})$. The r.h.side of the second equation in (\ref{eq:139.2}) for $d \delta_k/dt$ is of the
order of 1. However in the next equations in (\ref{eq:139.2}) there enter only the differences $\delta_k - \delta_{k\pm 1}$ whose
derivative, taking into account eq. (\ref{eq:7'}) is again $ \mathcal{O}(\varepsilon^{1/2})$. The right hand sides of the other
two equations in (\ref{eq:139.2}) are of the order of $ \mathcal{O}(\varepsilon)$. 

In addition we need also the evolution equations for the polarization vectors $\vec{n}_k$ which are:
\begin{equation}\label{eq:nk}\begin{split}
i \frac{\partial \vec{n}_k }{ \partial t} +8\nu _0^2 \sum_{n=k\pm 1}^{} e^{-\Delta _{k,n}}R_{kn} \vec{n}_n
  +C\vec{n}_k =0,
\end{split}\end{equation}
where $C$ is a $k$-independent real constant.

Thus for the unperturbed case $\alpha_2=0$ we have proven that $\frac{\partial \vec{n}_k}{ \partial t}
=\mathcal{O}(\epsilon)$. Since the scalar products $\langle\vec{n}_k^\dag , \vec{n}_n\rangle$ appear in the right hand side of eq. (\ref{eq:139.2})
in the factors  $e^{-\Delta_{kn}}R_{kn}$, that are itself of the order $\mathcal{O}(\epsilon)$ it is evident that
we can neglect the evolution of $\vec{n}_k$ using the initial values of $\left. \langle\vec{n}_k^\dag , \vec{n}_n\rangle\right|_{t=0}$.

\subsection{Evolution of polarization vectors $\vec{n}_k$ -- effect of cross-modulation}

The cross-modulation changes substantially the evolution of the polarization vectors $\vec{n}_k$.
Indeed, inserting into
Eq.(\ref{eq:Lagr}) $\vec{n}_k^\dag$ instead of $u_k^*$
and neglecting the terms of the order of $\mathcal{O}(\epsilon)$ we obtain the following evolution equations for $\vec{n}_k$:
\begin{equation}\label{eq:nkt0}\begin{aligned}
i \frac{\partial n_{1k}}{ \partial t } &= \frac{ 2\alpha_2}{3} \left( |n_{1k}|^2 - |n_{2k} |^2 \right) n_{1k}, \\
i \frac{\partial n_{2k}}{ \partial t }&= - \frac{ 2\alpha_2}{3} \left( |n_{1k}|^2 - |n_{2k} |^2 \right) n_{2k}, ,
\end{aligned}\end{equation}
Divide the first line in (\ref{eq:nkt0}) by $n_{1k}$ and the second one by  $n_{2k}$. The imaginary parts give immediately
that
\begin{equation}\label{eq:nkt1}\begin{split}
 \frac{\partial |n_{1k}|}{ \partial t} = 0, \qquad  \frac{\partial |n_{2k}|}{ \partial t} = 0.
\end{split}\end{equation}
If we use the parametrization: (\ref{eq:nk0}) then
Eq.(\ref{eq:nkt1}) means that
\begin{equation}\label{eq:thetk}\begin{split}
 \frac{\partial \theta_k}{ \partial t } =0.
\end{split}\end{equation}
The  real parts give immediately that
\begin{equation}\label{eq:betak}\begin{split}
\beta_k(t) =  - \frac{2\alpha_2 }{3} \cos(2\theta_k) t  +\const.
\end{split}\end{equation}
Let us now calculate the scalar product:
\begin{equation}\label{eq:ndagn1}\begin{split}
\langle\vec{n}_{k+1}, \vec{n}_k\rangle =  \cos (\beta_{k+1} -\beta_k) \cos(\theta_{k+1}-\theta_k) - i  \sin (\beta_{k+1} -\beta_k) \cos(\theta_{k+1}+\theta_k).
 \end{split}\end{equation}
Now we can calculate the module and the phase of the scalar product:
\begin{equation}\label{eq:12}\begin{split}
|\langle\vec{n}_{k+1}, \vec{n}_k\rangle|^2 &= \cos^2 (\beta_{k+1} -\beta_k) \cos^2(\theta_{k+1}-\theta_k) +
  \sin^2 (\beta_{k+1} -\beta_k) \cos^2(\theta_{k+1}+\theta_k), \\
\arg \langle\vec{n}_{k+1}, \vec{n}_k\rangle &= -\arctan \left( \tan (\beta_{k+1} -\beta_k)
\frac{ \cos(\theta_{k+1}+\theta_k)}{\cos(\theta_{k+1}-\theta_k)} \right) .
\end{split}\end{equation}
Thus,  from Eqs.(\ref{eq:139.2})  we get:
\begin{equation}\label{eq:141.1}
{d(\mu _k+i\nu _k)  \over dt } = 4\nu _0 \left[
\langle\vec{n}_{k}, \vec{n}_{k-1}\rangle e^{q_{k}-q_{k-1}} -
\langle\vec{n}_{k+1}, \vec{n}_k\rangle e^{q_{k+1}-q_{k}} \right],
\end{equation}
where
\begin{equation}\label{eq:q_k}\begin{split}
q_k &= -2\nu _0\xi_k + k \ln 4\nu _0^2 - i (\delta _k+\delta _0 + k\pi -2\mu _0 \xi_k), \\
\nu _0 &= {1 \over N } \sum_{s=1}^{N} \nu _s, \qquad \mu _0 = {1 \over N } \sum_{s=1}^{N} \mu _s, \qquad
\delta _0 = {1 \over N } \sum_{s=1}^{N} \delta _s.
\end{split}\end{equation}
Besides, from (\ref{eq:139.2}) and (\ref{eq:q_k}) there follows (see \cite{PRE*55}):
\begin{equation}\label{eq:141.2}
{dq_k  \over dt } =-4\nu _0 (\mu _k + i\nu _k).
\end{equation}
and
\begin{equation}\label{eq:141.3}
{d^2q _k \over dt^2 } = 16\nu _0^2 \left[ \langle\vec{n}_{k+1}, \vec{n}_k\rangle e^{q_{k+1}-q_{k}} - \langle\vec{n}_{k},
\vec{n}_{k-1}\rangle e^{q_{k}-q_{k-1}} \right],
\end{equation}
which  proves the statement in \cite{Thes}. Eq.(\ref{eq:141.3}), combined with the system of equations for the
polarization vectors (\ref{eq:nkt0}) provides the proper generalization of the CTC  for the cross-modulated MM.
Note, that for $\alpha_2=0$ the scalar products $\langle\vec{n}_{k+1}, \vec{n}_k\rangle$  are time-independent and as a result
the GCTC (\ref{eq:141.3}) is integrable. For $\alpha_2\neq 0$ however, the scalar products $\langle\vec{n}_{k+1}, \vec{n}_k\rangle$
depend explicitly on $t$ as shown above, which makes the GCTC non-integrable.

The equations for the polarization vectors are nonlinear. So the
whole system of equations for $q_k$ and $\vec{n}_k$ seems to be rather complicated and
non-integrable even for the unperturbed MM. However, all terms in the right hand sides of the
evolution equations for $\vec{n}_k$ are of the order of
$\epsilon$. This allows us to neglect the evolution of $\vec{n}_k$ and to approximate them with their
initial values. As a result we obtain that the $N$-soliton
interactions for the multicomponent MM in the adiabatic approximation are
modeled by the CTC, see Section 3.

\section{GCTC and the Asymptotic Regimes of $N$-soliton Trains}

The fact that the $N$-soliton trains for the scalar nonlinear Schr\"odinger equation
are modeled by an integrable model -- GCTC allowed one to
predict their asymptotic behavior. The method to do so was based on the
exact integrability of the CTC \cite{PRL} and on its Lax representation.

Here we shall show, that similar results hold true also for the CTC (\ref{eq:139.2})
modeling the soliton trains of the MM.
Indeed, following Moser~\cite{Moser} we introduce the Lax pair
\begin{equation}\label{eq:K.3}
\dot{L} = [B, L],
\end{equation}
where
\begin{equation}\label{eq:LB}\begin{split}
L = \sum_{k=1}^N \left( b_k E_{kk} + a_k (E_{k,k+1} + E_{k-1,k}) \right), \qquad
B = \sum_{k=1}^N a_k\left( (E_{k,k+1} - E_{k-1,k} \right).
\end{split}\end{equation}
Here the matrices $(E_{kn})_{pq} = \delta_{kp}\delta_{nq} $, and
$E_{kn} = 0 $ whenever one of the indices becomes 0 or $N+1 $; the other
notations in (\ref{eq:K.3}) are as follows:
\begin{eqnarray}\label{eq:K.5}
a_k = {1 \over 2} \sqrt{\langle\vec{n}_{k+1}, \vec{n}_k\rangle} e^{(q_{k+1} - q_k)/2}, \qquad
b_k =  {1 \over 2} \left( \mu_k + i \nu_k \right).
\end{eqnarray}
One can check that the compatibility condition eq. (\ref{eq:K.3}) with $L$ and $B$ as in
(\ref{eq:LB}) is equivalent to the unperturbed GCTC (\ref{eq:139.2}).

The first consequence of the Lax representation is that the CTC has $N$ complex-valued integrals of motion
provided by the eigenvalues of $L$ which we denote by  $\zeta_k =\kappa_k+i\eta_k$, $k=1,\dots,N$. Indeed the Lax equation means that
the evolution of $L$ is isospectral, \emph{i.e.}, $d \zeta_k /dt =0$.

Another important consequence from the results of Moser \cite{Moser} is that for the real Toda chain
one can write down explicitly its solutions in terms of the scattering data, which consist of pair series $\{\zeta_k, r_k\}_{k=1}^{N} $ where
$r_k$ are the first components of the properly normalized eigenvectors of $L_0 $ \cite{Moser,Toda}.
For the real Toda chain both $\zeta_k=\kappa_k$ and $r_k$ are real; besides all $\zeta_k$ are different.
Next, Moser  calculated the asymptotics of these solutions for $t\to\pm\infty$ and showed that $\kappa_k$
determine the asymptotic velocities of the particles.

The formulae derived by Moser can easily be extended to  the complex case \cite{JPA1}.
The important difference is that all important ingredients such as  eigenvalues
$\zeta_k$ and first components of the eigenvectors of $L$ normalized to 1 now become
complex-valued. In addition, the important asset of $L$ for the RTC, namely that
all eigenvalues are real and different, is  also lost.
However the asymptotics of the solutions for $t\to\pm\infty$ can be calculated with the
result:
\begin{equation}\label{eq:L.1}
q_k(t) = -2\nu_0 \zeta_{k} t - B_k + {\cal  O} (e^{-Dt}) ,
\end{equation}
where $D $ in (\ref{eq:L.1}) is some real positive constant, which is estimated by the
minimal difference between the asymptotic velocities.
Equating the real parts in Eq.~(\ref{eq:L.1}) we obtain:
\begin{equation}\label{eq:xi-k}
\lim_{t\to\infty} (\xi_k +2\kappa_k t ) = \const,
\end{equation}
which means that the real parts $\kappa_k$ of the eigenvalues of $L$ determine the asymptotic
velocities  for the CTC.
This fact can be used to classify the regimes of asymptotic behavior.

Let us also remind the important differences between RTC and CTC. 
The only possible asymptotic behavior in the RTC is the asymptotically free motion of the soliton.
For CTC the variety is much more depending on the number of solitons in the trains we differ three asymptotic regimes:
asymptotically free regime, bound state regime, and mixed asymptotic regimes (for details see, for example \cite{GTK-1,GT-3}).

The perturbed CTC taking into account the effects of the cross-modulation to the best of our knowledge
is not integrable and does not allow Lax representation. Therefore we are applying numeric methods to solve it.

 \section{Results and Discussion}
\begin{figure}[h!]
\centerline{\includegraphics[width=0.45\textwidth]{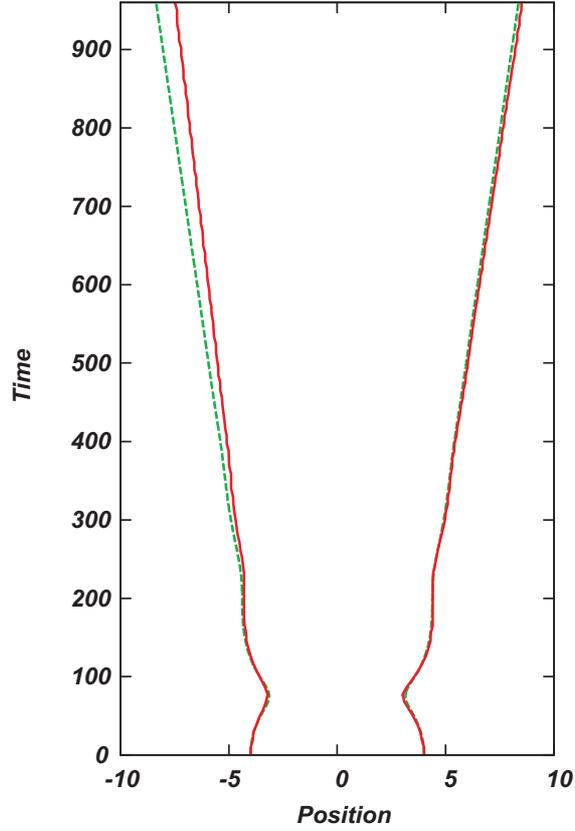}}
\vspace{-0.1in}
\caption{$\alpha_2=0$: $\delta_{1,2}=0,\pi$; $\nu_{1,2}=0.5$; $\xi_{1,2}=\pm4$; $\mu_{1,2}=0$; $\beta_{1,2}=\pi/3, -\pi/4$;
$\theta_1=\pi/4$, $\theta_2=\theta_1-\pi/20$. MM solution is plotted by solid line. CTC solution -- by dashed line.} 
\label{fig1}
\end{figure}

\begin{figure}[h!]
\centerline{\includegraphics[width=0.85\textwidth]{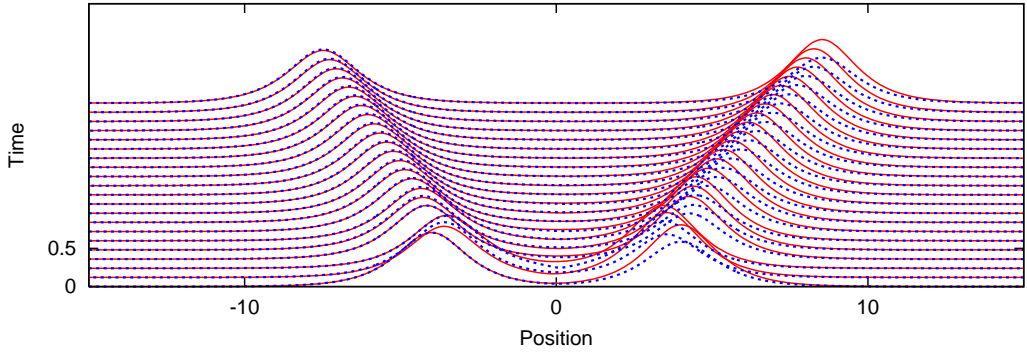}}
\vspace{-0.1in}
\caption{MM solution corresponding to data from Figure \ref{fig1}.}
\label{fig1a}
\end{figure}

In order to understand better the influence of the cross-modulation we will reduce our considerations to two-soliton trains.
We start with the case of trivial cross-modulation, \emph{i.e.}, $\alpha_2=0$. In this case the equations \eqref{eq:Man1} become MM.
In this case the trajectories of the soliton centers predicted by CTC and finite-difference implementation of MM are in excellent agreement. (Figure \ref{fig1}). The asymptotical behavior is asymptotically free regime. The temporal behavior of the MM solution is plotted in a separate graph (Figure \ref{fig1a}).

\begin{figure}[h!]
\centerline{\includegraphics[width=0.45\textwidth]{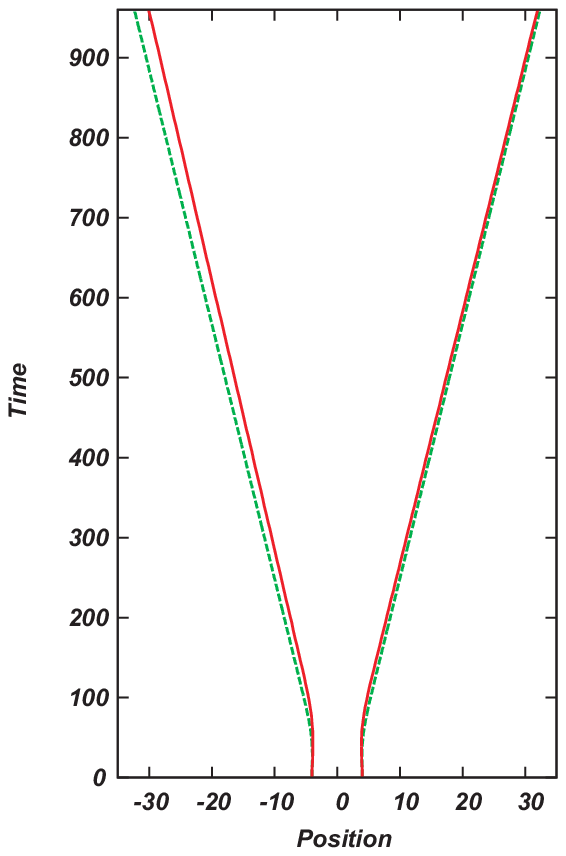}\includegraphics[width=0.45\textwidth]{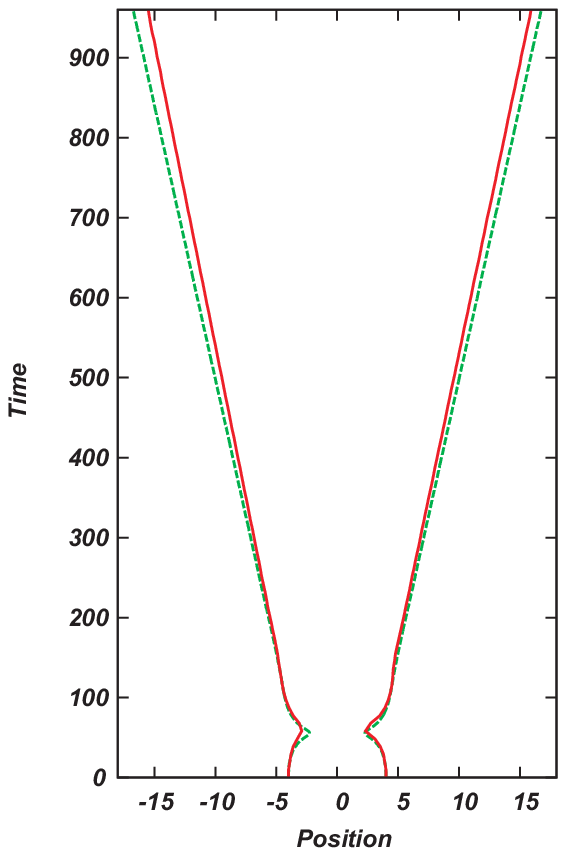}}
\vspace{-0.1in}
\caption{$\delta_{1,2}=0,\pi$; $\nu_{1,2}=0.5$; $\xi_{1,2}=\pm4$; $\mu_{1,2}=0$; $\beta_{1,2}=\pi/3, -\pi/4$; $\theta_1=\pi/4$,
$\theta_2=\theta_1-\pi/20$; $\alpha_2=-0.045$: (left);  $\alpha_2=0.045$ (right).  MM solution is plotted by solid line. CTC solution -- by dashed line.}
\label{fig2}
\end{figure}

\begin{figure}[h!]
\centerline{\includegraphics[width=0.45\textwidth]{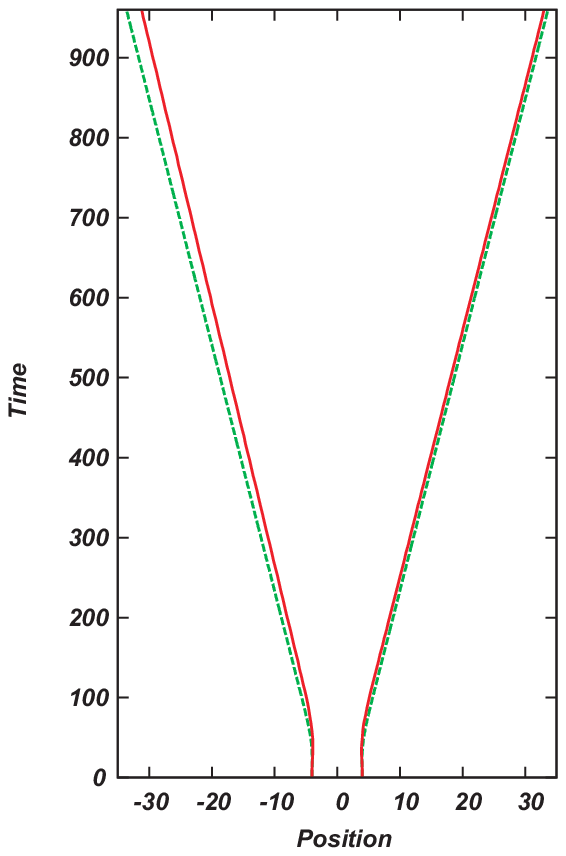}\includegraphics[width=0.45\textwidth]{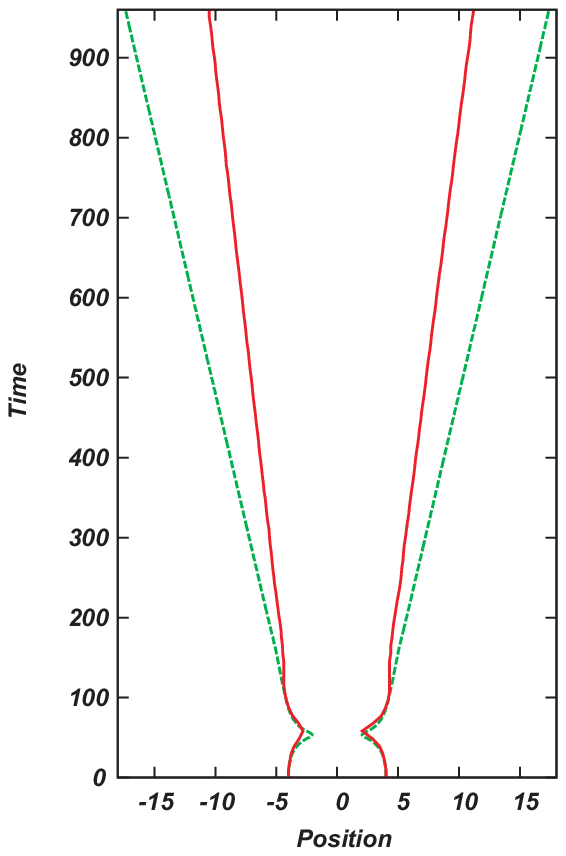}}
\vspace{-0.1in}
\caption{$\delta_{1,2}=0,\pi$; $\nu_{1,2}=0.5$; $\xi_{1,2}=\pm4$; $\mu_{1,2}=0$; $\beta_{1,2}=\pi/3, -\pi/4$;
$\theta_1=\pi/4$, $\theta_2=\theta_1-\pi/20$; $\alpha_2=-0.06$: (left);  $\alpha_2=0.06$ (right). MM solution is plotted by solid line. CTC solution -- by dashed line.}
\label{fig3}
\end{figure}

\begin{figure}[h!]
\centerline{\includegraphics[width=0.45\textwidth]{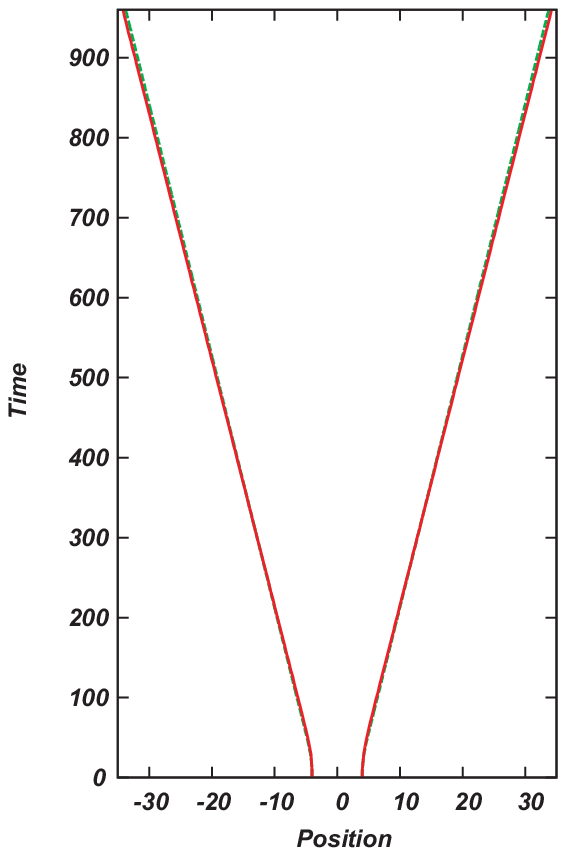}\includegraphics[width=0.45\textwidth]{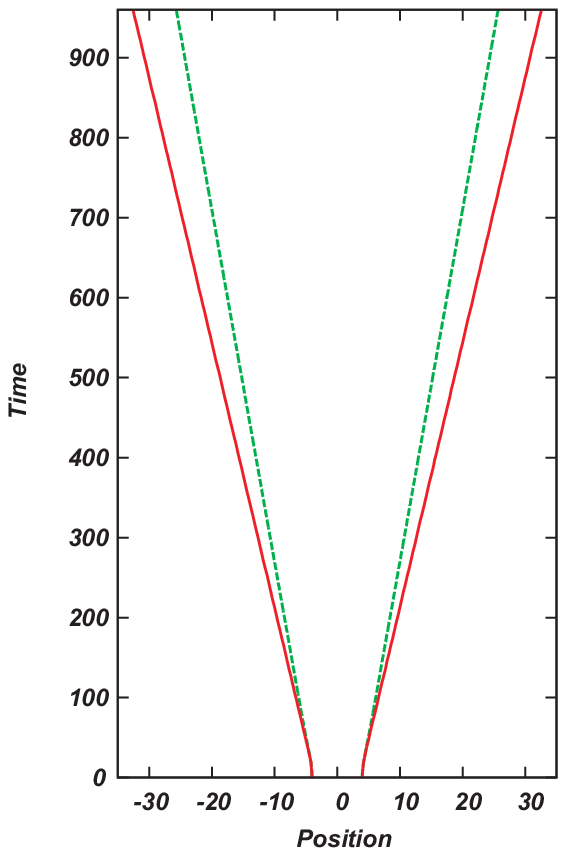}}
\vspace{-0.1in}
\caption{$\alpha_2=0.06$: $\delta_{1,2}=0$; $\nu_{1,2}=0.5$; $\xi_{1,2}=\pm4$; $\mu_{1,2}=0$; $\beta_{1,2}=\pi/3, -\pi/4$;
 $\theta_{1,2}=\pi/4\pm\pi/20$ (left); $\delta_{1,2}=0,\pi$; $\nu_{1,2}=0.5$; $\xi_{1,2}=\pm4$; $\mu_{1,2}=0$; $\beta_{1,2}=0$;
 $\theta_{1,2}=\pi/4\pm\pi/20$  (right). MM solution is plotted by solid line. CTC solution -- by dashed line.}
\label{fig4}
\end{figure}

Having in mind the multiparametric manifold we fix the parameters in Figure \ref{fig1} and vary only the cross-modulation. In Figure \ref{fig2} the cross-modulation $\alpha_2=-0.045$ (left) and $\alpha_2=0.045$ (right), respectively. In both cases the asymptotic behavior
is conserved though a phase shift of the trajectories of the soliton centers is observed.
One more pair of computations with the same set of initial parameters are conducted, this time for $\alpha_2=\pm0.06$. The results
are plotted in Figures \ref{fig3}-left and \ref{fig3}-right and they are not differ qualitatively from those in  Figures \ref{fig2}.
From the next pair of plots it is seen that the asymptotic prediction of CTC in presence of non-trivial cross-modulation
can be improved if one varies the values of the initial phases
$\delta_{1,2}$ (Figure \ref{fig4}-left) and polarization angles $\beta_{1,2}$  (Figure \ref{fig4}-right).
In all the considered cases angles $\theta_{1,2}$ are in the vicinity of $\pi/4$.

\begin{figure}[h!]
\centerline{\includegraphics[width=0.85\textwidth]{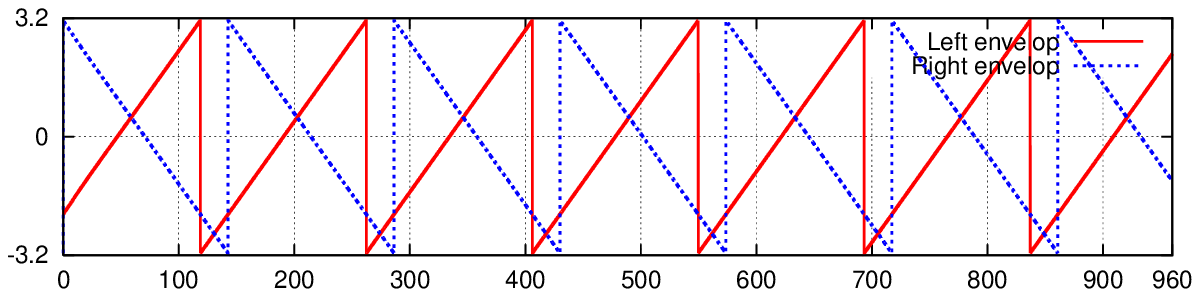}}
\vspace{-0.1in}
\caption{$\alpha_2=0.06$: Linear behavior of the polarization angles  $\beta_{1,2}$ from Figure \ref{fig4}-left.}
\label{fig5}
\end{figure}

\begin{figure}[h!]
\centerline{\includegraphics[width=0.85\textwidth]{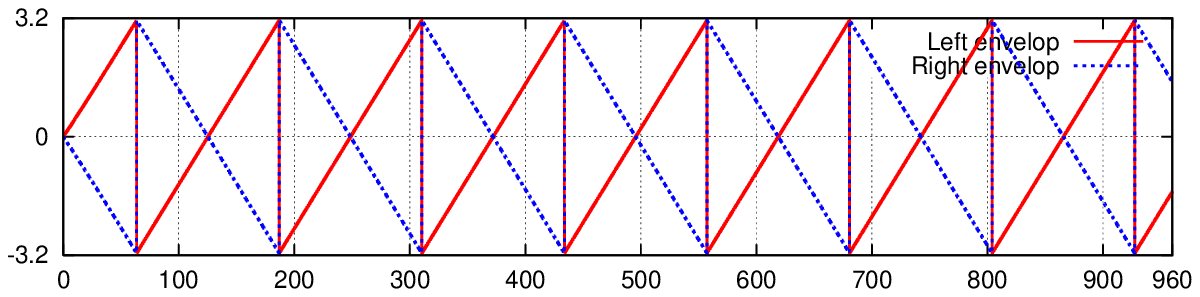}}
\vspace{-0.1in}
\caption{$\alpha_2=0.06$: Linear behavior of the polarization angles  $\beta_{1,2}$ from Figure \ref{fig4}-right.}
\label{fig6}
\end{figure}

\begin{figure}[h!]
\centerline{\includegraphics[width=0.85\textwidth]{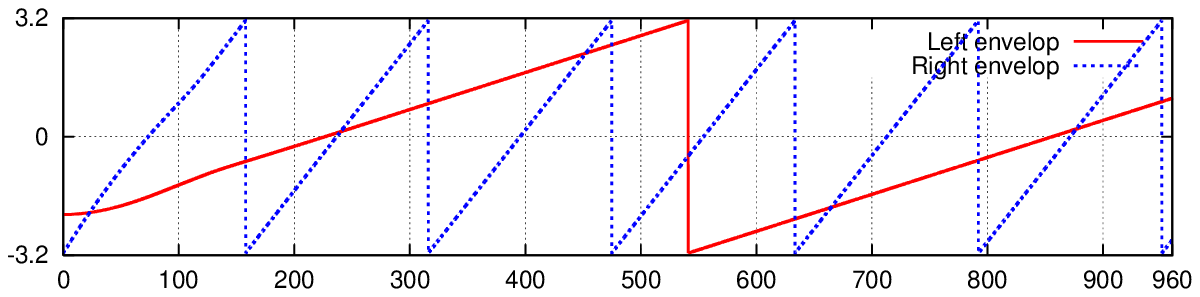}}
\vspace{-0.1in}
\caption{$\alpha_2=-0.06$: Linear behavior of the polarization angles  $\beta_{1,2}$ from Figure \ref{fig3}-left.}
\label{fig7}
\end{figure}

Next, the second our result is related to the linear temporal behavior of the polarization angles $\beta_{1,2}$. Concerning
the last Figure \ref{fig4}-left GCTC gives slopes of the left soliton envelope $0.0123$ and for the right soliton envelope -- $-0.024$.
To make sure in the validity of these numerical predictions, Eq.~\eqref{eq:nkt1}, we evaluate the polarization angles from MM in another way, \emph{i.e.},
$$4\beta_{1,2}=\Im {\rm Log} \frac{u_1 u_2^*}{u_1^*u_2}$$ where values of functions $u_1$ and $u_2$
correspond to the maxima of their envelopes, \emph{i.e.}, centers of the solitons. The latter formula comes right after equations \eqref{eq:Nstr} and \eqref{eq:nk0}. For the slopes we get $\pm 0.011$.
The linear behavior of the polarization angles based on the above formula is well visible in Figure \ref{fig5}.
GCTC gives the same predictions for the slopes of the polarization angles of the solitons plotted in Figure
\ref{fig4}-right, while MM predicts $\pm0.127$. The linear behavior is presented in Figure \ref{fig6}.
We consider and compare one more case, when $\alpha_2=-0.06$ (Figure \ref{fig3}-left). GCTC predicts zeroth slope
of $\beta_1$ for the left soliton envelop and $0.021$ -- for the right one. MM predicts $0.0002$ and $0.01$, respectively.
The temporal behavior of $4\beta_{1,2}$ is given in Figure \ref{fig7}. The angles have been evaluated and plotted by module $2\pi$. These
comparisons confirm cogently the GCTC prediction, Eq.\eqref{eq:nkt1}, for the linear evolution of the polarization angles $\beta_{1,2}$.

\section*{Conclusions}
It is well known  that strong cross-modulation ( i.e. large $\alpha_2$)   is not an adiabatic process.
In fact it leads to creation of additional quasi-particles  during the collisions, see Ref. \cite{CT08}. 

That is why we studied the influence of weak cross-modulation (i.e. small $\alpha_2$) on the soliton trains of MM in the adiabatic approximation.  We showed that for small values
of $\alpha_2 \simeq \sqrt{\epsilon}$  the adiabatic method still holds true. In particular, the GCTC model predicts that the
angles $\beta_k$ parametrising the polarization vectors (\eqref{eq:nk0}) depend linearly on time,  which is confirmed by the numerical
results, see Figures 5--7. The investigations elucidate the transition of the soliton solutions of the integrable Manakov model to the non-integrable nonlinear vector Schr\"odinger equation.

\section*{Acknowledgements}
We thank anonymous referees for their criticism and for  careful reading the text  of the paper
which lead to its improvement. 
This investigation is partially supported by Bulgarian National Science Fund under grant I-02/9.

\appendix

\section{Typical integrals}
Here we list the typical integrals that appear in calculating the action for the cross-modulated MM.
More details about their derivation and use to treat the effects of external potentials are given in
\cite{GT-3} and the references therein.

We start with Fourier integrals of hyperbolic functions \cite{GraRy}:
\begin{equation}\label{eq:JJp}
{\cal  J}_p(a) = \int_{-\infty}^{\infty} {dz \, e^{iaz} \over 2 \cosh^p z} = \int_{0}^{\infty} {dx\, \cos ax \over \cosh^p x }
,\qquad J_p(a) = \int_{-\infty}^{\infty} {dz \, e^{iaz} \sinh z \over 2 \cosh^p z}.
\end{equation}
They satisfy the recurrent relations:
\begin{equation}\label{eq:JPP-2}\begin{aligned}
{\cal J}_p(a)  &= {a^2 + (p-2)^2 \over (p-1) (p-2) }{\cal J}_{p-2}(a) , &\qquad J_{p}(a) &= {ia \over p-1 } {\cal J}_{p-1}(a) ,\\
{\cal   J}_{1}(a)  &= {\pi  \over 2\cosh {a\pi \over 2 } }, &\qquad {\cal   J}_{2}(a)  &= {\pi a \over 2\sinh {a\pi \over 2 } } .
 \end{aligned}\end{equation}
Next we need more complicated integrals \cite{GraRy}:
\begin{equation}\label{eq:K}\begin{split}
K(a,\Delta )&\equiv  \int_{-\infty }^{\infty } {dz\;e^{iaz}  \over 2\cosh z \cosh (z+\Delta ) } = {\pi \over  \sinh \Delta} {\sin
(a\Delta/2) \over \sinh (\pi a/2) } e^{-ia\Delta /2} = {\pi (1-e^{-ia\Delta }) \over 2i \sinh(\Delta )\sinh(\pi a/2) }, \\
L(a,\Delta )&\equiv  \int_{-\infty }^{\infty } {dz\;e^{iaz}\sinh z \over 2\cosh^2 z \cosh (z+\Delta ) } = {\pi i \over 2\sinh^2 \Delta \sinh (\pi
a/2) } \left[ (1-e^{-ia\Delta }) \cosh \Delta  - ia \sinh \Delta \right],
\end{split}\end{equation}

Of course, in our calculations we need only the terms of order $\varepsilon \simeq |\Delta| e^{-|\Delta|}$:
\begin{equation}\label{eq:K0}\begin{split}
K(0,\Delta )&\equiv  \int_{-\infty }^{\infty } {dz  \over 2\cosh z \cosh (z+\Delta ) } = {\Delta  \over \sinh \Delta } , \\
L(0,\Delta )&\equiv  \int_{-\infty }^{\infty } {dz\;\sinh z \over 2\cosh^2 z \cosh (z+\Delta ) } = {1 \over \sinh \Delta }
\left[ 1- {\Delta \cosh \Delta \over \sinh \Delta }\right],
\end{split}\end{equation}
which easily follow from the above expressions.

\section*{References}


\end{document}